\documentclass[10pt,a4paper]{article}                                  

\usepackage[T1]{fontenc}
\usepackage[utf8]{inputenc}
\usepackage[english]{babel}
\usepackage[]{graphics}
\usepackage{wrapfig}
\usepackage{caption}
\usepackage{amsmath}

\usepackage{amssymb}
\usepackage{graphicx}
\usepackage[export]{adjustbox}
\usepackage{float}
\usepackage{subfig}
\usepackage{color}
\usepackage[usenames,dvipsnames]{xcolor}
\usepackage{pdflscape}
\usepackage{cite}
\usepackage{graphics} 
\usepackage{epsfig} 
\usepackage{mathptmx} 
\usepackage{amsmath} 
\usepackage{amssymb}  

\newtheorem{remark}{Remark}
\newtheorem{assumption}{Assumption}

\newtheorem{theorem}{Theorem}
\newtheorem{lemma}{Lemma}
\newtheorem{proof}{Proof}[section]

\begin{document}
	
	\date{}
	
	\title{On multi-step prediction models for receding horizon control}
	\author{Enrico Terzi, Lorenzo Fagiano, Marcello Farina, Riccardo Scattolini \thanks{The authors are with the Dipartimento di Elettronica, Informazione e Bioingegneria, Politecnico di Milano, Milano, Italy. E-mail addresses:
			{\tt\small \{enrico.terzi| lorenzo.fagiano |marcello.farina| riccardo.scattolini\}@polimi.it }}}

\maketitle

\section{Introduction}\label{S:intro}
This manuscript contains technical details of recent results developed by the authors on learning-based model predictive control for  linear time invariant systems.

\section{Problem formulation}\label{Statement}

Let us consider a single-input, single-output (SISO), open-loop stable, discrete-time, strictly proper linear time invariant (LTI) system with $n$ states, input $u(k)\in\mathbb{R}$ and output $z(k)\in\mathbb{R}$, where $k\in\mathbb{Z}$ is the discrete time variable. The measurement of the system output $y(k)$ is affected by a bounded additive disturbance $d(k)$:
\begin{equation}\label{E:meas_output}
y(k)=z(k)+d(k).
\end{equation}
\begin{assumption}\label{dbound} (Measurements)
	\begin{itemize}
		\item $|d(k)|\leq\overline{d},\,\forall k\in\mathbb{Z}$.
		\item The input $u(k)$ is measured with negligible noise.\hfill$\square$
	\end{itemize} 
\end{assumption}
\vspace{0.2cm}
\begin{assumption}\label{ubound} (Input bounds) 
The input $u$ belongs to a compact set: $u(k) \in \mathbb{U}\subset\mathbb{R}, \forall k\in \mathbb{Z}$. \hfill$\square$
\end{assumption}
\vspace{0.2cm}
Let us denote with $p\in\mathbb{N}$ a finite number of time steps in the future. We are interested in deriving a prediction model of the future output $z(k+p)$, exploiting the input and output measurements collected in the time interval $[k-o+1,k]$, where $o\in\mathbb{N}$ is the chosen order of the model, and the future (planned) inputs in the interval $[k,k+p-1]$.  Specifically, at any instant $k$ let us define the regressor $\varphi_p(k)\in \mathbb{R}^{2o-1+p}$ as: 
\begin{equation}
\varphi_p(k)\doteq\left[ Y_o^T(k) \quad U_o^T(k) \quad U^T_p(k) \right]^T
\label{regressor}
\end{equation}
where $^T$ is the matrix transpose operation and
\[
\begin{split}
Y_o(k) & \doteq\left[
\begin{array}{c}
y(k) \\
y(k-1) \\
\vdots \\
y(k-o+1) \\
\end{array}
\right]               
U_o(k)\doteq\left[
\begin{array}{c}
u(k-1) \\
\vdots \\
u(k-o+1) \\
\end{array}
\right] \\
U_p(k) & \doteq\left[
\begin{array}{c}
u(k) \\
u(k+1)\\
\vdots \\
u(k+p-1) \\
\end{array}
\right].
\end{split}
\]
Then, we consider the following linear model structure:
\begin{equation}
\hat{z}(k+p)=\varphi_p(k)^T\theta_p
\label{E:model_structure}
\end{equation}
where $\hat{z}(k+p)$ is the predicted output and $\theta_p \in \mathbb{R}^{2o-1+p}$ is the model parameter vector. We refer to models of the form \eqref{E:model_structure} as ``multi-step'', since for each $p\in\mathbb{N}$ the corresponding prediction model provides directly an estimate of $z(k+p)$. This is different from the (most common) approach of deriving a one-step-ahead model and then iterating it $p$ times to compute predictions for the subsequent future time-steps.\\
In addition to the prediction model, we also want to derive guaranteed bounds on its accuracy. In particular, we aim for the following \emph{global} guaranteed accuracy bound:
\begin{equation}
|z(k+p)-\hat{z}(k+p)|\leq\tau_p(\theta_p).
\label{E:uncertainty_bounds}
\end{equation}
The value $\tau_p(\theta_p)$ is termed ``global'' since it holds for any value of the regressor $\varphi_p$ within a specified set, as we further detail in the remainder of this paper.\\
A collection of multi-step models derived for all $p\in[1,\overline{p}]$ provides an estimated sequence of future system outputs, up to the prediction horizon $\overline{p}<\infty$, together with an associated sequence of guaranteed uncertainty intervals $\tau_p(\theta_p)$. These models can be then embedded in a robust finite-horizon optimal control problem to be solved in a receding-horizon approach, thus realizing a MPC law based on multi-step predictions \cite{TFFS18b}. Three important reasons to consider multi-step models are:
\begin{enumerate}
	\item the model identification procedure results in convex optimization problems, as opposed to the nonlinear programs arising when identifying one-step-ahead models with an output-error (i.e. simulation) criterion \cite{Ljung99};
	\item the guaranteed bound $\tau_p(\theta_p)$ pertaining to a multi-step model is less conservative than the one pertaining to a one-step-ahead model iterated $p$ times;
	\item by considering an independent model for each value of $p$, there is no need for trade-offs between model accuracy at high-frequency (i.e. short prediction horizon) vs. low-frequency (i.e. long prediction horizon) that arise when choosing the simulation horizon in the identification of one-step-ahead models with an output-error criterion \cite{shook1991identification}. 
\end{enumerate}
The use of a model structure that is linear in the parameters is justified both by point 1) above and by the observation that, in the case of zero measurement noise, the true output $z(k+p)$ is indeed a linear function of the regressor, provided that the following assumption holds.
\vspace{0.2cm}
\begin{assumption}\label{A:model_order} (Observability, reachability and model order) 
	The system at hand is completely observable and reachable, and $o \geq n$. \hfill$\square$
\end{assumption}
\vspace{0.2cm}
Under Assumption \ref{A:model_order}, with straightforward manipulations one can show that the state at time $k$ is in general a linear function of the past $n$ input-output values, and that the output at time $k+p$ is linear in the state and in the planned input values, hence resulting in the model structure \eqref{E:model_structure}. From the practical standpoint, Assumption \ref{A:model_order} can be relaxed to account only for the observable and controllable sub-space of the system state. Moreover, this assumption can be satisfied by estimating the system order $n$ (e.g. based on physical considerations) and/or by deriving models with growing order $o$ and by monitoring the magnitude of the corresponding accuracy bounds, as we describe more in detail in section \ref{SS:tuning}.

Assumption \ref{ubound}, combined with the asymptotic stability of the system under study, results in bounded sets where the regressor $\varphi_p$ evolves in time. Specifically, for a given horizon $p$ we consider a compact set $\Phi_p$ containing the regressor values of interest:
\[
\varphi_p(k)\in\Phi_p\subset{R}^{2o-1+p},\,\text{$\Phi_p$ compact},\,\forall p\in\mathbb{N},\,\forall k\in\mathbb{Z}.
\]
$\Phi_p$ needs not to be known explicitly and is in general a complicated set that depends on the system input-output trajectories. Rather, we assume that for a given value of $p$ a finite batch of experimental data is available:
\begin{equation}\label{E:data_batch}
\left(\tilde{\varphi}_p(i),\,\tilde{y}_p(i)\right),\,i=1,\ldots,N_p.
\end{equation}
In \eqref{E:data_batch}, the notation $\tilde{\cdot}$ indicates specific measured values of a quantity. $\tilde{\varphi}_p(i)$ are the measured instances of the regressor $\varphi_p\in\Phi_p$, and  $\tilde{y}_p{(i)}$ the corresponding measured values of noise-corrupted outputs $p$ steps in the future, i.e. $\tilde{y}_p{(i)}=z(i+p)+d(i+p)$.  $N_p<\infty$ is the total number of available pairs $\left(\tilde{\varphi}_p(i),\,\tilde{y}_p(i)\right)$ composing the data-set. These pairs can be easily built from a given data-set of measured input-output values, collected e.g. in a preliminary experiment on the system. We further define the vectors of sampled variables $\tilde{v}_p(i)\doteq[\tilde{\varphi}_p(i)^T\;\tilde{y}_p(i)]^T,\,i=1,\ldots,N_p$ and the corresponding countable set containing them:
\begin{equation}\label{E:data_batch2}
\tilde{\mathcal{V}}_p^{N_p}=\left\{\tilde{v}_p(i) = \left[\begin{array}{c} \tilde{\varphi}_p(i)\\\tilde{y}_p(i)\end{array} \right],\, i=1,\dots N_p\right\}\subset\mathbb{R}^{2o+p}.
\end{equation}
For any given value of $\varphi_p\in\Phi_p$, the corresponding measured system output $y_p$ is not uniquely determined a priori, rather it belongs to a set $\boldsymbol{Y}_p(\varphi_p)\subset{R}$, due to the measurement noise $d$ (both in the regressor and in the corresponding output). In view of Assumptions \ref{dbound} and \ref{A:model_order} and of the compactness of $\Phi_p$, the set $\boldsymbol{Y}_p(\varphi_p)$ is also compact for any $\varphi_p\in\Phi_p$. We can then define the following continuous counterpart of the set $\tilde{\mathcal{V}}_p^{N_p}$:
\begin{equation}\label{E:data_set_continuous}
\mathcal{V}_p=\left\{v_p \doteq \left[\begin{array}{c} \varphi_p\\y_p\end{array} \right]: y_p\in\boldsymbol{Y}_p(\varphi_p), \forall \varphi_p\in\Phi_p\right\}\subset\mathbb{R}^{2o+p}.
\end{equation}
Namely, the set $\mathcal{V}_p$ contains all possible regressors $\varphi_p$ in the compact $\Phi_p$ and, for each value of  $\varphi_p$ , all possible output values in the corresponding compact set  $\boldsymbol{Y}_p(\varphi_p)$. We consider the following assumption linking the sets $\tilde{\mathcal{V}}_p^{N_p}$ and $\mathcal{V}_p$.
\begin{assumption}\label{A:data_set} (Data-set) 
	For any $\beta>0$, there exists a value of $N_p<\infty$ such that:
	\[
	d_2\left(\mathcal{V}_p,\tilde{\mathcal{V}}_p^{N_p}\right)\leq\beta
	\]
	where $d_2\left(\mathcal{V}_p,\tilde{\mathcal{V}}_p^{N_p}\right)\doteq\max\limits_{v\in\mathcal{V}_p}\min\limits_{u\in\tilde{\mathcal{V}}_p^{N_p}}\|u-v\|_2$ is the distance between sets $\mathcal{V}_p$ and $\tilde{\mathcal{V}}_p^{N_p}$. \hfill$\square$
\end{assumption}
Assumption \ref{A:data_set} is equivalent to assuming that $\lim\limits_{{N_p} \to \infty}d_2\left(\mathcal{V}_p,\tilde{\mathcal{V}}_p^{N_p}\right)=0$, i.e. that by adding more points to the measured data set, the underlying set of all trajectories of interest is densely covered. This is essentially an assumption on the persistence of excitation of the inputs used for the preliminary experiments, together with an assumption of bound-exploring property of the additive disturbance $d$, such that the bound $\overline{d}$ in Assumption \ref{dbound} is actually tight. 

We are now in position to formulate the problem that we will address in the remainder of this paper:\hfill
\vspace{0.2cm}\\
\textbf{Problem 1} Under Assumptions \ref{dbound}-\ref{A:data_set}, for a given $\overline{p}<\infty,\,p\in\mathbb{N}$ and any $p=1,\ldots,\overline{p}$, use the available data \eqref{E:data_batch} to identify a multi-step model of the form \eqref{E:model_structure} and estimate the associated guaranteed bounds \eqref{E:uncertainty_bounds}.\hfill$\square$
\vspace{0.2cm}\\
We propose in the next section an approach to solve \textbf{Problem~1}, based on a Set Membership (SM) identification methodology \cite{MNPW96}, \cite{Traub80}, which guarantees convergence of the derived error bounds to suitably defined optimal values.

\section{Learning multi-step prediction models:\\a set membership approach} \label{Identification}

For the sake of simplicity, in the following we consider a single value of $p\in[1,\overline{p}]$, the extension to any other value is straightforward. The proposed approach consists of the following steps.
\begin{enumerate}
	\item Define an optimality criterion to evaluate the model estimates, corresponding to an optimal (i.e. minimal) error bound.
	\item Derive a procedure to estimate the optimal error bound.
	\item Based on the available data and the error bound estimate, build the set of all parameters that are consistent with this information (Feasible Parameter Set, FPS).
	\item Using the information summarized in the FPS, for any given model of the form \eqref{E:model_structure} compute the related guaranteed error bound $\tau_p$.
	\item Select a nominal model with minimal guaranteed error bound.
\end{enumerate}
We describe next each step in detail, followed by a discussion on tuning and extensions of the approach.

\subsection{Optimal parameters and optimal error bound}\label{SS:optim_def}

Assume that a value of $\theta_p$ has been fixed. Then, under the considered working assumptions, $\forall k\in\mathbb{Z}$:
\begin{equation} \begin{split}
y(k+p)&=z(k+p)+d(k+p)\\
&=\hat{z}(k+p)+\epsilon_p(\theta_p,\varphi_p(k))+d(k+p)\\
&=\varphi_p(k)^T\theta_p+\epsilon_p(\theta_p,\varphi_p(k))+d(k+p)
\label{hyp1}
\end{split} \end{equation}
where $\epsilon_p(\theta_p,\varphi_p(k))$ is the error between the true system output and the estimated one:
\begin{equation} 
\epsilon_p(\theta_p,\varphi_p(k))=z(k+p)-\varphi_p(k)^T\theta_p
\label{epsilon_1}
\end{equation}
The quantity $\epsilon_p(\theta_p,\varphi_p(k))$ accounts for the quality of the chosen parameter values, for the model order mismatch ($o>n$, compare Assumption \ref{A:model_order}) and for the noise in the regressor measurements. Since the underlying system dynamics are time-invariant,  $\epsilon_p$ depends inherently on the model parameter vector $\theta_p$ and on the specific  regressor $\varphi_p(k)$. $\epsilon_p(\theta_p,\varphi_p(k))$ is bounded because both  $y(k+p)$ and $\theta_p^T\varphi_p(k)$ are, due to the stability of the system and compactness of the set containing the input values. From \eqref{epsilon_1} and Assumption \ref{dbound} we have:
\begin{equation} \begin{array}{rl}
|y(k+p)-\varphi_p(k)^T\theta_p|&\leq |\epsilon_p(\theta_p,\varphi_p(k))|+\overline{d}\\
&\leq \bar{\epsilon}_p(\theta_p)+\overline{d}
\end{array}
\label{E:epsilon_bar}
\end{equation}
where $\bar{\epsilon}_p(\theta_p)$ is the global error bound with respect to all possible regressors of interest in the compact $\Phi_p$:
\begin{equation}\label{epsilon0}
\begin{array}{rl}\bar{\epsilon}_p(\theta_p)=&\min\limits_{\epsilon\in\mathbb{R}}\;\epsilon\\
&\text{subject to}\\
&|y_p-\varphi_p^T\theta_p|\leq\epsilon+\overline{d},\,\forall (\varphi_p,y_p):\left[\begin{array}{c} \varphi_p\\y_p\end{array} \right]\in\mathcal{V}_p
\end{array}
\end{equation}
The quantity $\bar{\epsilon}_p(\theta_p)$ is the tightest bound on the global (i.e. worst-case) estimation error that a given parameter vector  $\theta_p$ can produce. We can now define the  optimal parameter values (i.e. optimal models) as those that minimize such a bound. As a technical assumption, we consider all the parameters within a compact set $\Omega\subset\mathbb{R}^{2o-1+p}$. $\Omega$ can take into account application-specific prior information on the model parameters or, if no such information is available, it can be chosen as a large-enough set (e.g. by considering box constraints of $\pm10^{15}$ on each element of the parameter vector). This assumption allows us to use maximum and minimum operators instead of supremum and infimum. The set $\Theta_p^0$ of optimal parameter values is the following:
\begin{equation}\label{theta^o}
\Theta_p^0=\left\{\theta_{p}^{0}:\,
\theta_{p}^{0}=\arg\min\limits_{\theta_p \in \Omega}\;\bar{\epsilon}_p(\theta_p)\right\},
\end{equation} 
and we denote with $\bar{\epsilon}_p^0$ the corresponding optimal error bound:
\begin{equation}\label{epsilon^o}
\bar{\epsilon}_p^0=\min\limits_{\theta_p \in \Omega}\;\bar{\epsilon}_p(\theta_p).
\end{equation} 
Considering \eqref{epsilon0}-\eqref{epsilon^o} we can alternatively write:
\begin{equation}\label{theta^o2}
\Theta_p^0=\left\{\theta_p : |y_p-\varphi_p^T\theta_p|\leq\bar{\epsilon}_p^0+\overline{d},\,\forall (\varphi_p,y_p):\left[\begin{array}{c} \varphi_p\\y_p\end{array} \right]\in\mathcal{V}_p\right\}.
\end{equation} 

\subsection{Estimating the optimal error bound}\label{SS:opt_estim}

The optimal models and optimal error bound cannot be computed in practice, since the solution to \eqref{theta^o} would imply the availability of an infinite number of data and the solution to an infinite-dimensional optimization program. However, we can compute an estimate $\underline{\lambda}_p\approx\bar{\epsilon}_p^0$ from the available experimental data, by solving the following linear program (LP):
\begin{equation}
\begin{array}{c}
\underline{\lambda}_p=\min\limits_{\theta_p,\lambda}\,\lambda\\
\text{subject to}\\
|\tilde{y}_p-\tilde{\varphi}_p^T\theta_p|\leq\lambda+\overline{d},\,\forall (\varphi_p,y):\left[\begin{array}{c} \tilde{\varphi}_p\\\tilde{y}_p\end{array} \right]\in\tilde{\mathcal{V}}_p^{N_p}\\
\lambda \geq 0
\end{array}
\label{E:bound_estim}
\end{equation}
The last inequality constraint in \eqref{E:bound_estim} is required to enforce a positive estimate of the error bound: without this constraint, the obtained estimate could result to be negative, especially in presence of small amount of data and output disturbance realizations with much smaller magnitude than the considered bound $\overline{d}$.

The following result shows that, under the considered working assumptions, the estimate \eqref{E:bound_estim} converges to the optimal one, $\bar{\epsilon}_p^0$.
\vspace{0.2cm}
\begin{theorem}
	Let Assumptions \ref{dbound}-\ref{A:data_set} hold. Then:
	\begin{enumerate}
		\item $\underline{\lambda}_p\leq\bar{\epsilon}_p^0$;
		\item $\forall \rho\in(0,\bar{\epsilon}_p^0]\; \exists \; N_p<\infty\; :\; \underline{\lambda}_p\geq\bar{\epsilon}_p^0-\rho$\hfill$\square$
	\end{enumerate}
	\label{convergence}
\end{theorem}
\vspace{0.2cm}
\begin{proof}
	See the appendix.
\end{proof}
Theorem \ref{convergence} implies that $\lim\limits_{N_p\to\infty}(\bar{\epsilon}_p^0-\underline{\lambda}_p)=0^+$, i.e. that the estimate \eqref{E:bound_estim} converges to the optimum from below. This is unavoidable with the considered problem settings, where limited information is available. In practice, one can increase the number $N_p$ of experimental data and observe the behavior of $\underline{\lambda}_p$, which typically (if the preliminary experiments are informative enough) quickly converges to a limit. Then, a practical approach to compensate for the uncertainty caused by the use of a finite number of measurements  is to inflate the value $\underline{\lambda}_p$:
\begin{equation}\label{E:alpha}
\hat{\bar{\epsilon}}_p=\alpha\underline{\lambda}_p,\;\alpha>1.
\end{equation}
With sufficiently large $N_p$, the coefficient $\alpha$ can be chosen very close to 1. We show an example of such a procedure in section \ref{S:results}, and provide more comments on the tuning of $\alpha$ in section \ref{SS:tuning}. We consider the following assumption in the remainder of this paper:
\vspace{0.2cm}
\begin{assumption} (Estimate of the optimal error bound)\label{A:estim_bound}\\
	$\hat{\bar{\epsilon}}_p\geq\bar{\epsilon}_p^0$.\hfill$\square$
\end{assumption}
\vspace{0.2cm}

\subsection{Feasible Parameter Set}\label{FeasibleParameterSet}

We can now exploit the estimated optimal error bound to construct the tightest set of parameter values that are consistent with all the prior information, i.e. the FPS $\Theta_p$:
\begin{equation}
\Theta_p=\left\{\theta_p : |\tilde{y}_p-\tilde{\varphi}_p^T\theta_p|\leq\hat{\bar{\epsilon}}_p+\overline{d},\,\forall (\tilde{\varphi}_p,\tilde{y}_p):\left[\begin{array}{c} \tilde{\varphi}_p\\\tilde{y}_p\end{array} \right]\in\tilde{\mathcal{V}}_p^{N_p}\right\}
\label{FPSdef}
\end{equation} 

The set $\Theta_p$ is non-empty by construction, since under Assumption \ref{A:estim_bound} we have (compare \eqref{theta^o2} and \eqref{FPSdef}):
\begin{equation}\label{E:FPS_contains_Theta0}
\Theta_p^0\subseteq\Theta_p.
\end{equation} 
If the FPS is bounded, it results in a polytope with at most $N_p$ faces. If it is unbounded, then this is a sign that the employed measured data are not informative enough to derive a bound on the worst-case model error (as we show next) and that the number of available data $N_p$ shall be increased until a bounded FPS is obtained. This situation usually occurs when very few data points are used (e.g. $N_p<2o-1+p$) or the preliminary experiments are not informative enough (e.g. when only steady-state data are used).

\subsection{Error bound computation for a generic model}\label{SS:error_bounds}
Having defined the FPS, we can proceed to derive a bound on the prediction error achieved by any  model of the form \eqref{E:model_structure}. Let us consider a generic value of $\theta_p$ to derive the prediction model \eqref{E:model_structure}. Then, considering any $\theta_p^0\in\Theta_p^0$ and any $\varphi_p(k)\in\Phi_p$, using \eqref{epsilon_1} and \eqref{theta^o2} we have:
\begin{equation} \begin{array}{rl}
&\lvert z(k+p) - \hat{z}(k+p) \rvert\\
=&\lvert \varphi_p(k)^T\theta_p^0 - \varphi_p(k)^T\theta_p + \epsilon(\theta_p^0,\varphi_p(k)) \rvert \\
\leq& \lvert \varphi_p(k)^T(\theta^0_p-\theta_p)\rvert+\bar{\epsilon}_p^0
\label{prediction error}
\end{array}
\end{equation}
The tightest bound we can derive on \eqref{prediction error} is based on the knowledge that $\theta^0_p\in\Theta_p$ (see \eqref{E:FPS_contains_Theta0}) and that $\bar{\epsilon}_p^0\leq\hat{\bar{\epsilon}}_p$ (Assumption \ref{A:estim_bound}):
\begin{equation}\label{prediction error2}
\lvert \varphi_p(k)^T(\theta^0_p-\theta_p)\rvert+\bar{\epsilon}_p^0\leq
\max\limits_{\theta \in\Theta_p}\lvert \varphi_p(k)^T(\theta-\theta_p) \rvert + \hat{\bar{\epsilon}}_p.
\end{equation}
The latter bound is the tightest \emph{local} error bound (i.e. valid for a given value of $\varphi_p$) for the model given by the considered parameter value $\theta_p$. As stated in \textbf{Problem 1}, for the sake of using the model within a robust MPC framework, we want to derive a global bound $\tau_p(\theta_p)$ holding $\forall \varphi_p\in\Phi_p$. Considering \eqref{prediction error}-\eqref{prediction error2} leads to:
\[\begin{array}{rl}
&\max\limits_{\varphi_p\in\Phi_p}\lvert z(k+p) - \hat{z}(k+p) \rvert\\
\leq&\max\limits_{\varphi_p\in\Phi_p}\max\limits_{\theta \in\Theta_p}\lvert \varphi_p^T(\theta-\theta_p) \rvert + \hat{\bar{\epsilon}}_p\\
=&\tau_p(\theta_p)
\end{array}
\]
Such a bound cannot be derived exactly with finite data under the considered assumptions, and its computation would be intractable also if the set $\Phi_p$ were known precisely (unless some additional assumption is made, e.g. polytopic set $\Phi_p$). However, we can approximate it by computing the maximum of \eqref{prediction error2} over the data-set $\tilde{\mathcal{V}}_p^{N_p}$:
\begin{equation}\label{tau_estim}
\underline{\tau}_p(\theta_p)=\max\limits_{i=1,\ldots,N_p}\max\limits_{\theta \in\Theta_p}\lvert \tilde{\varphi}_p(i)^T(\theta-\theta_p) \rvert + \hat{\bar{\epsilon}}_p\\
\end{equation}
The following result shows convergence of $\underline{\tau}_p(\theta_p)$ to $\tau_p(\theta_p)$.
\begin{lemma}
	Let Assumptions \ref{dbound}-\ref{A:data_set} hold. Then, for any $\theta_p\in\mathbb{R}^{2o-1+p}$:
\begin{enumerate}
	\item $\underline{\tau}_p(\theta_p)\leq\tau_p(\theta_p)$;
	\item $\forall \rho\in(0,\tau_p(\theta_p)]\; \exists \; N_p<\infty\; :\; \underline{\tau}_p(\theta_p)\geq\tau_p(\theta_p)-\rho$\hfill$\square$
\end{enumerate}
\label{convergence2}
\end{lemma}
\vspace{0.2cm}
\begin{proof}
Straightforward extension of the proof of Theorem \ref{convergence}.
\end{proof}
Considerations similar to those of Theorem \ref{convergence} hold also for the bound $\underline{\tau}_p(\theta_p)$: it is possible to monitor its behavior with increasing $N_p$ value in order to evaluate its convergence. As done in \eqref{E:alpha}, we inflate this bound to account for the uncertainty deriving from our finite data-set:
\begin{equation}\label{E:gamma}
\hat{\tau}_p(\theta_p)=\gamma\underline{\tau}_p(\theta_p),\;\gamma>1,
\end{equation}
and we assume that the resulting estimate is larger than the true bound:
\begin{assumption} (Error bound estimate for a given $\theta_p$)\label{A:estim_bound2}\\
	$\hat{\tau}_p(\theta_p)\geq\tau_p(\theta_p),\,\forall\theta_p\in\mathbb{R}^{2o-1+p}$.\hfill$\square$
\end{assumption}

\subsection{Nominal model selection}\label{SS:nominal model}
The last step in the proposed approach is to select a nominal model. The most common approach is probably least-squares estimation, in which case the results of section \eqref{SS:error_bounds} can be anyway applied to obtain an estimate of the resulting global error bound. Since the final goal is to employ the model in a MPC algorithm, we rather seek the model that minimizes the uncertainty bound. Specifically, considering that the tightest set that contains the optimal parameter values (i.e. with minimum error, see section \eqref{SS:optim_def}) is the FPS $\Theta_p$, we search within this set for the parameter value that minimizes the resulting bound $\hat{\tau}_p(\theta_p)$:
\begin{equation}
\theta_p^*=\arg\min\limits_{\theta_p\in\Theta_p}\hat{\tau}_p(\theta_p).
\label{minmaxmax_orig}
\end{equation}
The resulting nominal model is $\hat{z}(k+p)=\varphi_p(k)^T\theta_p^*$, and the associated error bound is:
\begin{equation}\label{E:nom_bound}
\hat{\tau}_p(\theta_p^*)=\gamma\left(\min\limits_{\theta_p\in\Theta_p}\max\limits_{i=1,\ldots,N_p}\max\limits_{\theta \in\Theta_p}\lvert \tilde{\varphi}_p(i)^T(\theta-\theta_p) \rvert\right) + \hat{\bar{\epsilon}}_p.
\end{equation}
Note that term $\hat{\bar{\epsilon}}_p$ in \eqref{E:nom_bound} does not depend on $\theta^*_p$ and it converges to the optimal error bound $\bar{\epsilon}_p^0$ as $N_p$ increases (Theorem \ref{convergence}). Moreover, it can be shown that, under the considered working assumptions, if $\theta_p^0$ is unique then the difference $|\theta_p^0-\theta_p^*|$ tends to zero, and the associated error bound \eqref{E:nom_bound} tends to the minimum value $\bar{\epsilon}_p^0$ as well. Furthermore, for any value of $N_p$ it can be shown that $\hat{\tau}_p(\theta_p^*)$ corresponds to the radius of information, i.e. the minimum guaranteed error value that can be attained with the given prior information and data \cite{Traub80}. Finally, it also holds that, for each $p>1$ the bounds $\hat{\tau}_p(\theta_p^*)$ computed in our approach are less conservative than the worst-case bounds obtained by iterating any 1-step-ahead prediction model of the form \eqref{E:model_structure}. All these derivations are omitted here for the sake of brevity. 
\begin{remark} \label{R:computation} The optimization problem to be solved to compute $\theta_p^*$ \eqref{minmaxmax_orig} takes the form 
\begin{equation}
\theta_p^*=\arg\min\limits_{\theta_p\in\Theta_p}
\max_{i=1,\ldots,N_p}\max\limits_{\theta \in\Theta_p}\lvert \tilde{\varphi}_p(i)^T(\theta-\theta_p) \rvert.
\label{minmaxmax}
\end{equation}
This problem can be solved by reformulating it as $2\,N_p+1$ LPs \cite{boyd2004convex}, as reported in the appendix for completeness. 
\end{remark}

\subsection{Learning algorithm, tuning aspects and extensions}\label{SS:tuning}

In the closing part of this section we report the overall identification algorithm, to be repeated for each prediction step $p=1\ldots \overline{p}$:
\begin{enumerate}
	\item Collect $N_p$ measured regressor values and the corresponding measured output instances (see \eqref{E:data_batch})\\
	\item Solve the optimization problem \eqref{E:bound_estim} and compute $\hat{\bar{\epsilon}}_p=\alpha\underline{\lambda}_p$, that is needed to define the FPS \eqref{FPSdef}\\
	\item Solve \eqref{minmaxmax} to derive the nominal model $\theta^*_p$ and select a value of $\gamma>1$ to compute the worst case prediction error estimate $\hat{\tau}_p(\theta_p^*)$ \eqref{E:nom_bound}.
\end{enumerate}
The main tuning parameters in the approach are the model order $o$, the disturbance bound $\overline{d}$, and the scalars $\alpha$ and $\gamma$. In practical applications, a good estimate or even exact knowledge of $o$ and $\overline{d}$ is often available from considerations on the plant and the available sensors. If not, an order selection procedure consists, as anticipated in section \ref{Statement}, in monitoring the estimate $\hat{\bar{\epsilon}}_p$ with increasing values of $o$: typically there is a clear convergence to a constant or slowly decreasing bound when $o\geq n$. We provide an example of this procedure in the next section. Regarding the choice of $\overline{d}$, over- or under-estimating this value leads,  respectively, to either a too optimistic estimate of the prediction errors (since part of the model uncertainty is then hidden in the disturbance bound) or to a higher conservativeness (since the prediction error bound is then accounting also in part for the measurement noise). In a similar way, too large values of $\alpha$ and $\gamma$ can lead to conservative error bounds, while values too close to 1 might give error bounds that are too tight and could be violated by new, previously unseen data. As a matter of fact, both conditions (over- and under-estimation) can be easily monitored on-line, and the derived tuning parameters $\overline{d},\,\alpha$ and $\gamma$ suitably adjusted. This feature can be exploited in view of employing the proposed approach in an adaptive framework, which is a planned extension of this work. Other planned extensions that can be developed relatively easily are to consider nonlinear systems, since we can still employ the same model parametrization and the described approach by embedding the effect of nonlinearities in the error term $\epsilon_p(\theta_p,\varphi_p)$ (see \eqref{hyp1}), and the use of other choices of linearly-parametrized models, e.g. truncated series of functional bases, since the same approach and main results would still hold, referred to the specific chosen class of models.

\section{Simulation results}\label{S:results}

The proposed algorithm has been tested on the benchmark example already considered in \cite{shook1991identification}. In particular, the data are generated according to the continuous-time transfer function:
\begin{equation}
G(s)=2\frac{229}{(s+1)(s^2+30s+229)}
\end{equation}

The input-output samples are collected with a sampling time $T_s = 0,2 s$ according to the output equation $y(k) = z(kT_s)+d(kT_s)$, where $d(t)$ is a colored noise obtained by low-pass filtering a randomly generated number, with a first-order filter with time constant of $0.2\,$s. The disturbance $d(t)$ is bounded in the interval $[-0.2\;0.2], \forall t$, and the prediction horizon is $\bar{p}=10$.

The collected dataset is made of 500 input-output data samples. The input is a three level signal taking values randomly each $4\,$s in the set $\{-1,0,1\}$.

\begin{figure}[thpb]
\centering
\includegraphics[width=1\linewidth]{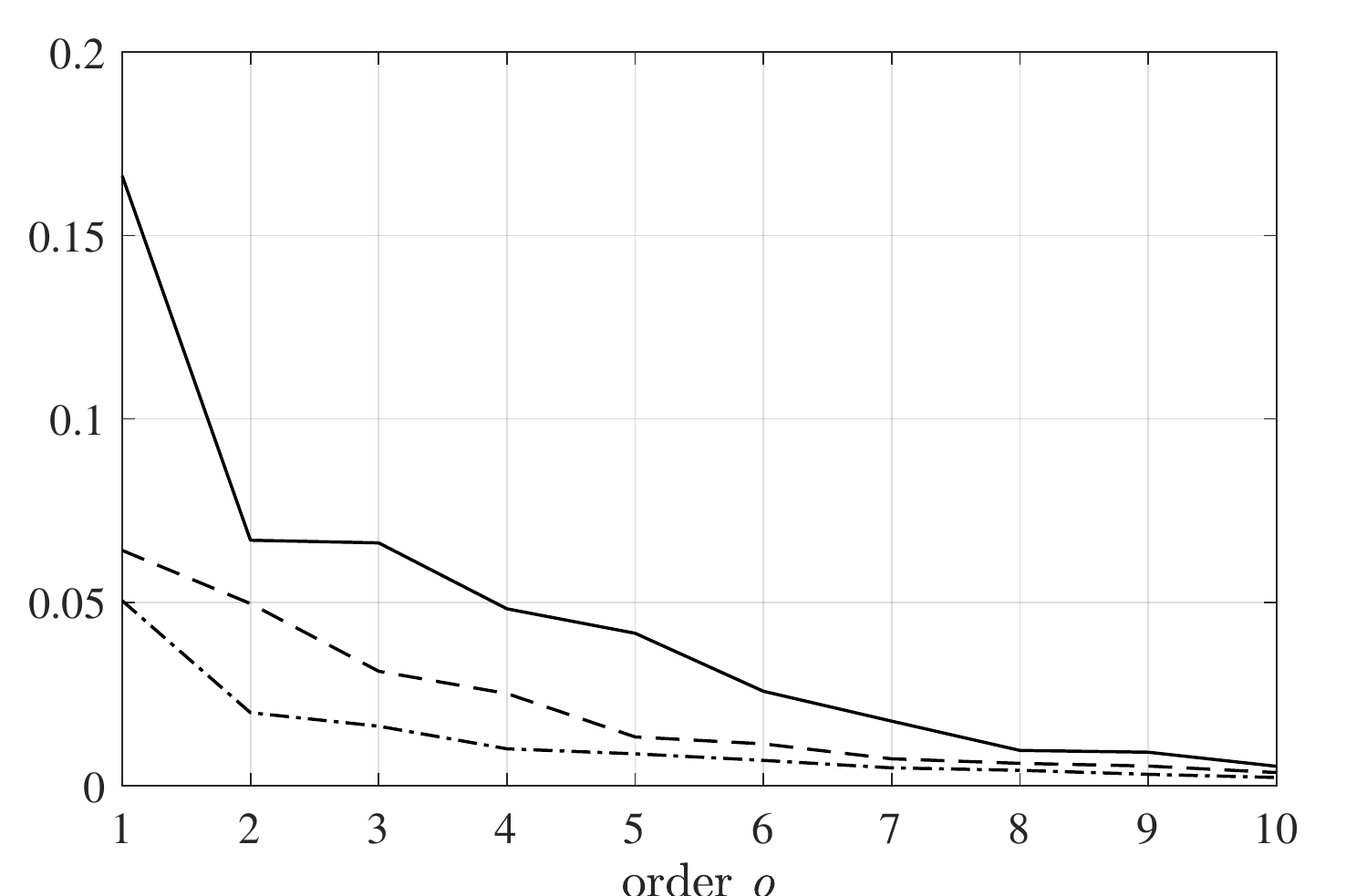}
\caption{Values of $\underline{\lambda}_p$ for different orders of the model. Solid lines: $p=3$; dashed black lines: $p=6$; dash-dot line: $p=9$ \label{fig:lambdaorder}}
\end{figure}

In order to select the order of the model, we solve problem \eqref{E:bound_estim} for different orders $o$, as reported in Fig. \ref{fig:lambdaorder}. The trend of the solution is monotonically decreasing, and this can be explained thanks to the noisy regressors that increase in number in vector $\varphi_p$ as $o$ grows (see \eqref{regressor}). The effect of noise contained in them vanishes over time due to compensation of terms, eventually reducing the bound $\underline{\lambda}_p \forall p=\dots \bar{p}$ for growing order $o$ of the model. The final choice has been $o=3$, consistently with Assumption \ref{A:model_order}, that matches the order of the system, even though its true dynamics are dominated by the pole in $\bar{s}=-1$. 

\begin{figure}[thpb]
\centering
\includegraphics[width=1\linewidth]{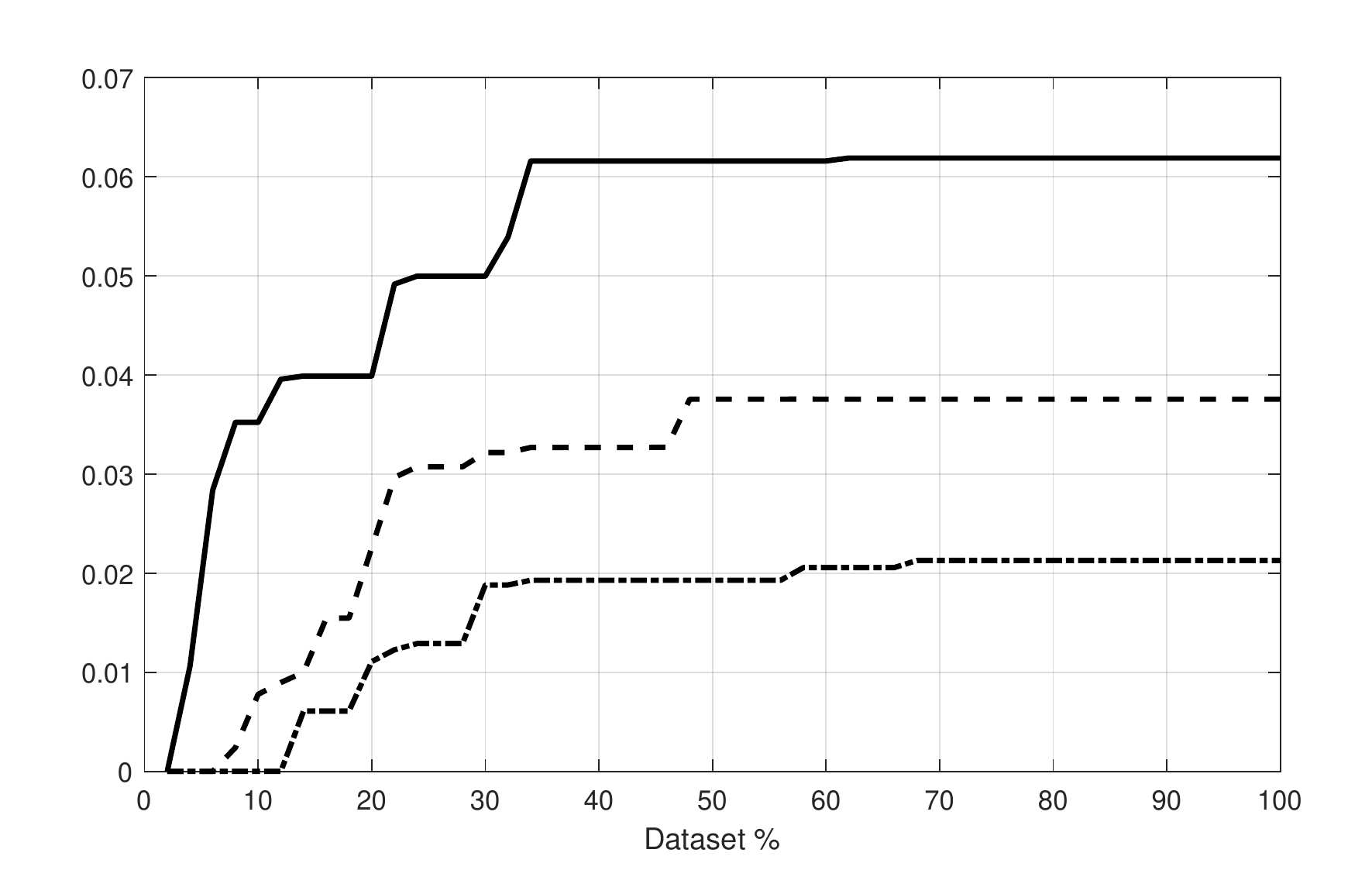}
\caption{Values of $\underline{\lambda}_p$ for different percentages of the dataset. Solid lines: $p=3$; dashed black lines: $p=6$; dash-dot line: $p=9$ } \label{fig:lambdadatap}
\end{figure}

Once $\underline{\lambda}_p$ is computed $\forall p=1 \dots \bar{p}$, it is inflated according to \eqref{E:alpha} with the fixed coefficient $\alpha=1.2$, i.e. with a $20\%$ additive margin. 
The conservativeness of such a choice can be evaluated from the trend that $\underline{\lambda}_p$ exhibits with respect to the number of data. In view of Theorem \ref{convergence}, in fact, $\lim\limits_{N_p\to\infty}(\bar{\epsilon}_p^0-\underline{\lambda}_p)=0^+$ and the convergence rate gives a qualitative idea of how reliable $\hat{\bar{\epsilon}}_p^0$ is with respect to the real unknown value $\epsilon_p^0$. Fig. \ref{fig:lambdadatap} presents the values of $\underline{\lambda}_p$ for the fixed steps $p=3,6,9$ with an increasing percentage of the dataset used for the computation. Using the whole dataset, the curves approach an asymptotic value which presumably correspond to the theoretical bound $\bar{\epsilon}_p^0$.

\begin{figure}[thpb]
\centering
\includegraphics[width=1\linewidth]{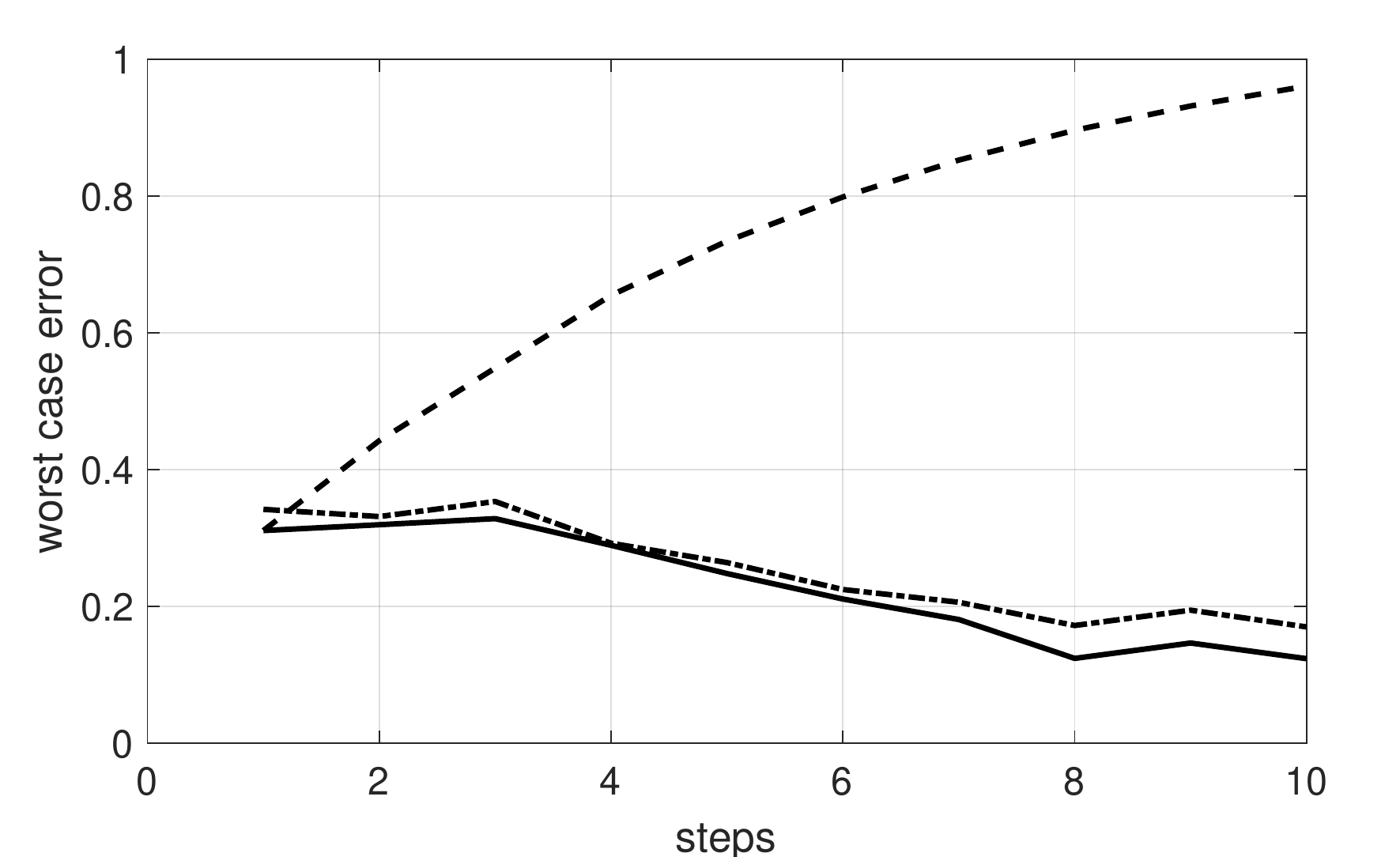}
\caption{Worst case error bounds. Solid lines: multistep approach ($\hat{\tau}_p(\theta_p^*)$); dashed black lines: iterated 1-step; dash-dot line: LS models $\hat{\tau}_p(\theta_p^{LS})$ } \label{fig:1multiLS}
\end{figure}

The multi-step predictors are eventually built by solving \eqref{minmaxmax} to compute the nominal model for each step and the related guaranteed worst-case error bound $\hat{\tau}_p(\theta^*_p)$. The resulting bounds are dramatically less conservative then those coming from the iteration of the 1-step-ahead predictor, as shown in Fig. \ref{fig:1multiLS}. 
We also carried out a comparison with the worst-case bounds associated to Least Square (LS) models, one for each $p$, reported in Fig. \ref{fig:1multiLS} too. The result  confirms the optimality of our proposed approach, that minimizes the uncertainty bounds. 
Note also that it may happen that a model computed with LS does not belong to the FPS. 
Finally, the descending trend of $\hat{\tau}_p(\theta^*_p)$ is due to the increasing number of noise-free input values $u(k)$ in the regressor as $p$ increases, combined with the fading effect of past input and output data.

\begin{figure}[thpb]
\centering
\includegraphics[width=1\linewidth]{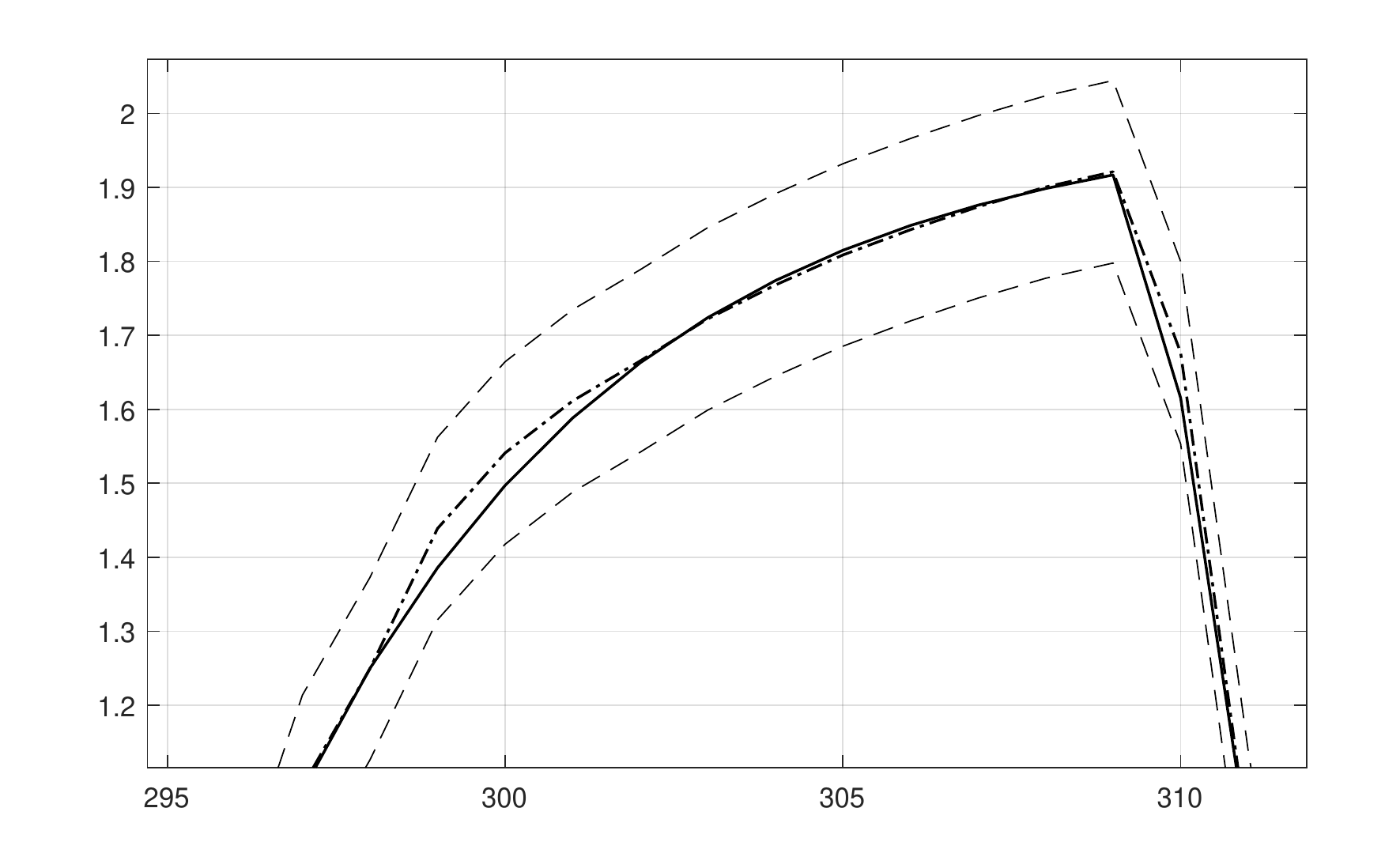}
\caption{Simulation with validation data for $p=\bar{p}=10$. Solid lines: real samples $z(k)$; dashed black lines: guaranteed bounds; dash-dot line: prediction with nominal model } \label{fig:simulation}
\end{figure}

The final outcome of the algorithm consists in $\bar{p}$ nominal models and as many guaranteed worst-case errors bounds, that are then tested with validation data. A detail of the simulation is reported in Fig. \ref{fig:simulation}.

\section{Conclusions}\label{S:conclusion}
We presented an approach to derive multi-step prediction models, and the related uncertainty bounds, for an unknown linear system affected by additive measurement disturbance. Under suitable assumptions, we demonstrated convergence of the bounds to their theoretical minimum based on the available information. The approach requires the solution to linear programs only. The derived models are particularly suited to robust model predictive control design, since they can predict the future trajectory of the system on a finite horizon and the related uncertainty intervals.

\section*{Appendix}

\subsection*{Proof of Theorem \ref{convergence}}

Problem \eqref{epsilon^o} can be rewritten as:
\begin{equation}
\begin{array}{ll}
&\min\limits_{\theta_p \in \Omega, \epsilon_p \in \mathbb{R}} \epsilon_p\\
&\text{subject to}\\
&|y_p-\varphi_p^T\theta_p|\leq\epsilon+\overline{d},\,\forall (\varphi_p,y_p):\left[\begin{array}{c} \varphi_p\\y_p\end{array} \right]\in\mathcal{V}_p
\end{array}
\label{eq:reformulation1}
\end{equation} 

At first, note that the solution to \eqref{eq:reformulation1} must imply a regressor, denoted with $\varphi_{p,0}$, and a corresponding output value, $y_{p,0}$, satisfying the constraint in (\ref{epsilon0}) with equality:
\begin{equation}
\bar{\epsilon}^o_p=\lvert y_{p,0}-\varphi_{p,0}^T\theta_p^0 \rvert - \bar{d} =\min\limits_{\theta_p\in\Omega} \lvert y_{p,0}-\varphi_{p,0}^T\theta_p  \rvert - \bar{d}
\label{epsilonbar0}
\end{equation}

\noindent\emph{Proof of claim 1)}. From the definition of $\mathcal{V}_p$ and $\tilde{\mathcal{V}}^{N_p}_p$ it holds that $\tilde{\mathcal{V}}^{N_p}_p \subset \mathcal{V}_p$; thus, from \eqref{E:bound_estim} neglecting the trivial case $\underline{\lambda}_p=0$, we have:
\begin{equation}
\begin{split}
\underline{\lambda}_p(N_p) = &\min\limits_{\theta_p \in \Omega}\max\limits_{\left[\begin{array}{c} \tilde{\varphi}_p\\
\tilde{y}_p\end{array} \right]\in\tilde{\mathcal{V}}_p^{N_p}} \lvert \tilde{y}_p - \tilde{\varphi}_p^T\theta_p \rvert - \bar{d} \leq \\
&\min\limits_{\theta_p \in \Omega}\max\limits_{\left[\begin{array}{c} \varphi_p\\y_p \end{array} \right]\in\mathcal{V}_p} \lvert \tilde{y}_p - \tilde{\varphi}_p^T\theta_p \rvert - \bar{d}
\label{eq:upperbound}
\end{split}
\end{equation}

In view of \eqref{epsilonbar0} we have then
\begin{equation}
\underline{\lambda}_p(N_p)  \leq \min\limits_{\theta_p \in \Omega}\lvert y_{p,0}- \varphi_{p,0}^T\theta_p  \rvert - \bar{d}
\end{equation}
which implies $\underline{\lambda}_p(N_p)  \leq \bar{\epsilon}_p^0 $.\\
$\,$\\
\noindent\emph{Proof of claim 2)}. Starting from \eqref{E:bound_estim}, with standard properties of absolute values, we compute
\begin{equation}
\begin{split}
\underline{\lambda}_p(N_p)=&\max\left\{0,\min\limits_{\theta_p \in\Omega} \max\limits_{\left[\begin{array}{c} \tilde{\varphi}_p\\
\tilde{y}_p\end{array} \right]\in\tilde{\mathcal{V}}_p^{N_p}}\lvert \tilde{y}_p-\tilde{\varphi}_p^{T} \theta_p\rvert - \bar{d}\right\} \geq \\
&\min\limits_{\theta_p \in\Omega}\left\{\lvert \bar{y}_p (N_p)-\bar{\varphi}_p(N_p)^T \theta_p\rvert - \bar{d} \right \}
\end{split}
\label{ineqlowerbound}
\end{equation}

where 
\begin{equation}
\begin{bmatrix} \bar{\varphi}_p(N_p) \\ \bar{y}_p(N_p)\end{bmatrix}= \arg\min\limits_{\left[\begin{array}{c} \tilde{\varphi}_p\\
\tilde{y}_p\end{array} \right]\in\tilde{\mathcal{V}}_p^{N_p}} \left\|\left[\begin{array}{c} \varphi_{p,0}\\
y_{p,0}\end{array} \right]-\left[\begin{array}{c} \tilde{\varphi}_p\\
\tilde{y}_p\end{array} \right] \right\|_2
\label{eq:mindist}
\end{equation}

Now adding and subtracting $y_{p,0}$ and $\varphi_{p,0}^T\theta_p$ from \eqref{ineqlowerbound} and neglecting the trivial case $\underline{\lambda}_p=0$:
\begin{equation}
\begin{split}
\underline{\lambda}_p(N_p) \geq &\min\limits_{\theta_p \in\Omega}\{\lvert (\bar{y}_p(N_p)-y_{p,0}) + \\
&(\varphi_{p,0}-\bar{\varphi}_p(N_p))^T\theta_p + y_{p,0} - \varphi_{p,0}^T\theta_p \rvert - \bar{d}\}\\
 \geq &\min\limits_{\theta_p \in\Omega}\{\lvert  y_{p,0} - \varphi_{p,0}^T\theta_p - (-\bar{y}_p(N_p)+y_{p,0}) + \\
&(\varphi_{p,0}-\bar{\varphi}_p(N_p))^T\theta_p \rvert - \bar{d}\} \\
 \geq&\min\limits_{\theta_p \in\Omega}\{\lvert  y_{p,0} - \varphi_{p,0}^T\theta_p \rvert \}-\\
&\min\limits_{\theta_p \in\Omega} \{\lvert (-\bar{y}_p(N_p) + y_{p,0}) +\\
& (-\varphi_{p,0}+\bar{\varphi}_p(N_p))^T\theta_p \rvert - \bar{d}\} \\
 =&\bar{\epsilon}_p^0 + \bar{d}-
\min\limits_{\theta_p \in\Omega} \{\lvert (-\bar{y}_p(N_p) + y_{p,0}) +\\
& (-\varphi_{p,0}+\bar{\varphi}_p(N_p))^T\theta_p \rvert - \bar{d}\} 
\end{split}
\end{equation}

Then, simplifying $\bar{d}$, we compute
\begin{equation}
\begin{split}
\underline{\lambda}_p(N_p) \geq \bar{\epsilon}_p^0 - \min\limits_{\theta_p \in\Omega} &\{\lvert (-\bar{y}_p(N_p) + y_{p,0}) \\
&+ (-\varphi_{p,0}+\bar{\varphi}_p(N_p))^T\theta_p \rvert\}
\end{split}
\end{equation}

Considering Assumption \ref{A:data_set}, we know that for $N_p \to \infty$
\begin{equation}
\forall \beta>0, \exists \bar{N}_p(\beta): \|\bar{\varphi}_{p}(\bar{N}_p)-\varphi_{p,0}\| \leq \beta, \lvert \bar{y}_{p}(\bar{N}_p)-y_{p,0} \rvert \leq \beta
\end{equation}

and thus, using the inequality $\lvert a^Tb\rvert \leq \|a\|_2\|b\|_2$
\begin{equation}
\begin{array}{ll}
\underline{\lambda}_p(\bar{N}_p(\beta)) &\geq\bar{\epsilon}_p^0 - \min\limits_{\theta_p \in\Omega}\{\lvert (-\bar{y}_p(\bar{N}_p(\beta)) + y_{p,0}) \rvert \\
&+\|\theta_p \|_2  \|\varphi_{p,0}-\bar{\varphi}_p(\bar{N}_p(\beta)) \|_2 \} \\
 &\geq\bar{\epsilon}_p^0 - \beta \left(1+ \sup\limits_{\theta_p \in \Theta_p^0} \|\theta_p\|_2 \right)
\label{eq:lowerbound}
\end{array}
\end{equation}

Finally, choosing $\beta \leq \frac{\rho}{\left(1+ \sup\limits_{\theta_p \in \Theta_p^0} \|\theta_p\|_2\right) } $ concludes the proof.

\subsection*{Reformulation of problem \eqref{minmaxmax}}
The problem \eqref{minmaxmax} can be stated as  \begin{equation}
\min\limits_{\theta_p\in\Theta_p}
\max_{i=1,\ldots,N_p}\max\limits_{\theta \in\Theta_p}\lvert \tilde{\varphi}_p(i)^T(\theta-\theta_p) \rvert
\label{eq:refoptprb}
\end{equation}

The absolute value in \eqref{eq:refoptprb} can be separated into two terms:
\begin{equation}
\min\limits_{\theta_p\in\Theta_p}\max\limits_{\theta \in\Theta_p}\max\limits_{i=1\dots N_p}\max \begin{pmatrix} \tilde{\varphi}_p(i)^T(\theta-\theta_p)\\	-\tilde{\varphi}_p(i)^T(\theta-\theta_p)\end{pmatrix}
\label{eq:refoptprb2}
\end{equation}

Let us now introduce the quantity $\check{\varphi}_p(j)=\begin{cases}
\varphi_p(i) \quad  if \quad  i\leq N_p\\
-\varphi_p(i) \quad if \quad i > N_p
\end{cases}$ and recast (\ref{eq:refoptprb2}) accordingly:
\begin{equation}
\begin{split}
&\min\limits_{\theta_p\in\Theta_p}\max\limits_{\theta \in\Theta_p}\max\limits_{j=1\dots 2N_p} \check{\varphi}_p(j)^T(\theta-\theta_p) = \\
&\min\limits_{\theta_p\in\Theta_p}\max\limits_{j=1\dots 2N_p}\max\limits_{\theta \in\Theta_p} \check{\varphi}_p(j)^T(\theta-\theta_p) = \\
&\min\limits_{\theta_p\in\Theta_p}\max\limits_{j=1\dots 2N_p}\max\limits_{\theta \in\Theta_p} \left(\check{\varphi}_p(j)^T\theta -\check{\varphi}_p(j)^T\theta_p\right)
\label{problem3}
\end{split}
\end{equation}
Let us define:
\[
c_j=\max\limits_{\theta \in\Theta_p} \check{\varphi}_p(j)^T\theta,\,j=1,\ldots,2N_p
\]
Then, \eqref{problem3} is equivalent to:
\begin{equation}
\min\limits_{\theta_p\in\Theta_p}\max\limits_{j=1\dots 2N_p} (c_j -\check{\varphi}_p(j)^T\theta_p)
\label{problemlastbutone}
\end{equation}

Problem (\ref{problemlastbutone}) corresponds to a single LP that reads:
\begin{equation}
\begin{split}
&\min\limits_{\theta_p\in\Theta_p} \zeta\\
s.t. \quad &c_j-\check{\varphi}_p(j)\theta_p \leq \zeta, \quad j=1\dots 2N_p\\
\end{split}
\end{equation}

\bibliographystyle{plain}

\end{document}